\documentclass[article]{elsarticle}
\usepackage{longtable}
\usepackage{aas_macros}
\usepackage{subfig}
\DeclareRobustCommand*\textsubscript[1]{%
  \@textsubscript{\selectfont#1}}
\def\@textsubscript#1{%
  {\m@th\ensuremath{_{\mbox{\fontsize\sf@size\z@#1}}}}}
\usepackage{bm}
\usepackage{amsmath,enumerate,url}
\usepackage{mathtools}
\usepackage{slashed}
\usepackage{amsfonts} 
\usepackage{MnSymbol} 
\usepackage{sidecap}
\usepackage{xcolor}
\usepackage[utf8]{inputenc}
\DeclareUnicodeCharacter{211D}{\mathbb{R}}
\journal{Journal of Radiation Physics and chemistry}
\usepackage{geometry} 
\usepackage{floatrow}
\usepackage{graphicx}
\usepackage{parskip} 

\begin{document}

\begin{frontmatter}

\title{Comprehensive calculations of energy levels, radiative transition parameters, hyperﬁne structure constants $A_J$ - $B_J$, Landé $g_J$ factors and isotope shifts for Sc XX}

\author[mymainaddress]{Shikha Rathi}

\author[mymainaddress]{Lalita Sharma\corref{mycorrespondingauthor}}
\cortext[mycorrespondingauthor]{Corresponding author}
\ead{lalita.sharma@ph.iitr.ac.in}

\address[mymainaddress]{Department of Physics, Indian Institute of Technology Roorkee, Roorkee 247667, India}

\begin{abstract}
Large scale calculations for the energy levels, transition rates, oscillator strengths, lifetimes, hyperfine interaction constants, Landé $g_J$ factors, and isotope shift factors have been performed for $1s^2$ and $1snl\hspace{1mm}(n=2-8$ and $l\leq n-1)$ levels of He-like Sc XX ion. The GRASP2018 package based on multi-configurational
Dirac–Fock method is used to obtain these results. The leading quantum electrodynamic corrections, Breit interaction and nuclear recoil effects are also included in the succeeding relativistic configuration interaction calculations. The relativistic isotope shift (RIS4) program is used to determine the mass and field shifts factors. A detailed comparison of the present results with the corresponding values from the NIST database and other theoretical work, wherever available, has been done, and an excellent agreement is achieved. A large section of the results is reported for the first time in the present work.
\end{abstract}

\begin{keyword}
\texttt{Energy levels; Transition parameters; Hyperfine interaction constants; Isotope shifts; He-like Scandium;}
\end{keyword}

\end{frontmatter}

\section{Introduction}
Atomic properties, e.g., energy levels and transition rates (A), are building blocks to interpret numerous processes in astrophysics, plasma physics, and other branches of science. Accurate knowledge of oscillator strengths (f), transition wavelengths etc. are essential for plasma modelling, determining the elemental composition of stars, calculating opacities and radiative accelerations. Nuclear effects like hyperfine interactions and isotope shifts are also important to investigate as they can change the line profile of the elements spectra \cite{wahlgren2005inputting}. 

Recently, scandium has received great attention due to its significant role in understanding stellar physics \cite{lawler2019transition}.
It is a transition metal, intermediate to $\alpha$ and iron-peak elements and has immense importance in astrophysics due to its presence in the solar photosphere \cite{prieto2020chemical, lodders2003solar, nissen2016high}, disc and metal rich halo stars \cite{nissen1999sc}, metal-poor stars \cite{sneden2016iron, cohen2004abundances}. 4.1 keV lines from the radioactive scandium ($^{44}$Sc) have been observed by the Chandra X-ray observatory \cite{weisskopf2000chandra} from G1.9+0.3 supernova remnant (SNR) \cite{reynolds2008youngest, borkowski2010radioactive}. Sc is a vital element to understand the ongoing processes in the AmFm stars \cite{leblanc2008scandium}. 
 Its unusual abundance in the galactic center nuclear star cluster provides a way to map their chemical composition \cite{do2018super}. Lines of scandium have also been observed in the exoplanet HD209458b within some uncertainties \cite{rojo2013ground}. Therefore, precision calculations for the atomic properties of Sc are required to help in reducing these uncertainties. It has been realized that a comprehensive atomic data of Sc is needed to resolve the contradiction between different abundance calculations and for effective implementation of the LTE and Non-LTE models for stellar atmosphere modeling, e.g., in metal poor stars \cite{zhang2008non}. Besides astrophysics,  it also has applications in the  laboratory plasma diagnosis where it can be used as an impurity e.g., in Alcator C-Mod tokamak \cite{rice1997impurity}. 
 
Scandium is usually present in its ionic form in the stars. Accurate electronic structures of its highly charged ions are pivotal to analyze star structure, and abundance \cite{1996A&A...310..872A}. He-like Sc XX is an essential ion out of its various ionic states, which can be found in broad temperatures ranges in astrophysical bodies and are responsible for the dominant features in their x-ray spectra. 
Spectra of Sc XX has also been observed in high-temperature fusion plasma from the Tokamak Alcator \cite{rice1995x}. To analyze these spectra, atomic properties, viz., energy levels, transition rates etc., are highly essential. Further, the hyperfine structure (HFS) calculations for Sc XX are also crucial as its nuclear spin is 7/2, which can lead to considerable splitting of energy levels due to hyperfine interactions. Several studies in the recent past have mentioned the importance of including hyperfine interactions to interpret spectra from astrophysical objects, and neglecting these effects can induce misleading abundance estimations \cite{jofre2015gaia}. Moreover, the detection of radioactive scandium necessitates isotope shift (IS) factors calculations. Such studies are crucial to rectify the isotope anomalies in astrophysical objects and  also play an important role in nuclear physics, e.g., to predict root mean square nuclear charge radii.  
Also, the presence of a magnetic field in stars accentuates the need for knowledge of Landé $g_J$-factors. \\
There are several theoretical studies reported on He-like Sc XX. Some of these are restricted to the states belonging to principal quantum number ($n$) equal to 2. These include the work of Singh et al. \cite{singh2020revised}, where in addition to energy levels and transition rates, they have also reported the hyperfine structure for low lying excited states pertaining to the configurations $1s2s$ and $1s2p$ of Sc XX ion using the multi-configuration Dirac-Fock (MCDF) method from the GRASP2k program \cite{jonsson2013new}. Plameri et al. \cite{palmeri2012atomic} used the Cowan's code \cite{cowan1981theory} to calculate electronic parameters for K-vacancy levels of Sc isonuclear sequence. 
Indelicato \cite{indelicato1988multiconfiguration} calculated the  energies required for transitions between $1s2p-1s2s$ and $1s2p-1s^2$ levels of helium iso-electronic series upto atomic number $Z=92$ using the MCDF method \cite{grant2007relativistic}. Qing et al. \cite{bo2008theoretical} implemented the MCDF method to compute the energy of the $2 ~^3P_{0,1,2}$ states of He-like ions upto $Z=36$. Drake \cite{drake1988theoretical} determined the energies of $1s^2, 1s2s$ and $1s2p$ levels of He-like ions till $Z= 100$ by using relativistic unified method. Employing the relativistic all-order many-body perturbation method, Plante et al. \cite{plante1994relativistic}  obtained the energies for $n=1$ and $n=2$ states of He-like ions for $Z=3-100$.
Higher $n$ values have been considered in the work of Natarajan and Natarajan \cite{natarajan2008kbeta} to obtain the energies of K$_\beta$ X-ray transitions from $1s3p-1s^2$ states of few He-like ions using the MCDF method. 
G{\l}owacki \cite{glowacki2020relativistic} computed the excitation energies and transition parameters for transitions among the $1s^2$, $1s2p$ and $1s3p$ states of He-like ions. Yerokhin and Surzhykov \cite{yerokhin2019theoretical} employed the relativistic configuration interaction (RCI) method and determined the energies of fine-structure states of $1sns$ and $1snp$ configurations with $n=3-7$ for He-like ions including Sc. Si et al. \cite{si2016energy} used the second-order many-body perturbation theory (MBPT) with flexible atomic code (FAC) \cite{gu2008flexible} to obtain the energies of $1s^2$ and $1snl(n\leq 6, l\leq(n-1))$ levels of Sc XX. They have also reported few transition rates for multipole transitions among these levels. Aggarwal et al. \cite{aggarwal2012energy} performed MCDF calculations to determine energies of Sc XX, for $1snl(n\leq5, l\leq(n-1))$ levels, using GRASP \cite{grant1980atomic} and FAC \cite{gu2008flexible}. Massacrier and Artru  \cite{massacrier2012extensive} considered various ionic stages of scandium (Sc III - SC XXI) and implemented RCI method using FAC \cite{gu2008flexible} to obtain the energy levels and  electric dipole transition probabilities for  $1s^2$ and $1snl(n\leq10, l\leq(n-1))$ and few double excited states. 
Bhatti et al. \cite{bhatti2001mcdf} used the MCDF method to determine the specific mass shift for the ground state of He-like ions. \\
Experimental works are scanty for He-like Sc XX ion. Boiko et al. \cite{boiko1978x} used the laser produced plasma technique and measured the wavelengths for $1s2p - 1s^2$ transitions. Beiersdorfer et al. \cite{beiersdorfer1989experimental}  performed wavelength measurements for $1snp$ $(n=3-5)$ to $1s^2$ transitions in Sc XX using the high-resolution PLT Johann spectrometer \cite{hill1979determination}. 
The NIST \cite{NIST} values for Sc XX were primarily from the work of Sugar and Corliss \cite{sugar1985atomic}, Kaufman and Sugar \cite{kaufman1988wavelengths} and Martin et al. \cite{martin1988atomic}. For modeling of stellar atmosphere, the data available at Kurucz data site \citep{Anu:2013} are generally used; however, for Sc XX, ion energies of only a few levels are listed there. \\
Thus, from the above literature review, we see that atomic structure results for He-like Sc XX are fragmented, and a complete analysis of Sc XX is required due to its utmost importance in various practical applications. Further, to the best of our knowledge, there are no reports on hyperfine structures, isotope shifts, and Landé g$_J$ factors for its highly excites states except for a few results from Singh et al. \cite{singh2020revised}, and Bhatti et al. \cite{bhatti2001mcdf}.
 These parameters are of prime importance, and their lack can scale down the efficacy of stellar atmosphere modeling and may lead to a fallacious prediction of elements' abundance in stars. \\
 In this regard, the objective of the present work is to carry out systematic large-scale calculations and provide accurate atomic-structure results for Sc XX. These calculations are performed using the GRASP2018 package \cite{fischer2019grasp2018} which employs a fully relativistic MCDF technique that accounts for the effect of electron correlations by considering many electronic configurations. Influence of the Breit interaction, specific mass shift corrections, vacuum polarization, and self-energy corrections are also included in the present work. We include the configurations $1snl (1 \leq n \leq 8$ and $0 \leq l \leq n-1 )$  in this study to determine the properties of the lowest 127 energy levels. In this connection, we report the transition parameters for electric and magnetic dipole (E1, M1) and quadrupole (E2, M2) transitions. Besides E1 transitions, the results for forbidden transitions are not only crucial in determining the precise lifetimes of energy levels but are also essential for effective and accurate plasma modeling. Further, we also calculate the lieftimes, hyperfine interaction constants, Landé $g_J$ factors, and isotope shifts for all the 127 levels to provide complete results. For IS calculations we employ the relativistic isotope shift (RIS4) module \cite{ekman2019ris} together with GRASP2018 \cite{fischer2019grasp2018}. Further, we provide a detailed comparison of our calculated energies, lifetimes, and transition parameters with the available theoretical and experimental results. A good agreement between different results assists in assessing the accuracy of the present HFSs, Landé $g_J$ factors, and  ISs calculations for which either no or extremely scarce results exist in literature to compare with. Therefore, most of the present results are completely new.\\
In the following Section 2, we briefly give the details of the theoretical method employed here. The computational procedure has been described in Section 3 where we discuss the generation of atomic state functions (ASFs). In Section 4, we have presented our results and their exhaustive comparison with the previously available results. 
\section{ Theory }

The GRASP2018 program \cite{fischer2019grasp2018}, used in the present work, is based on the MCDF method. In this approach, the Dirac–Coulomb Hamiltonian for an N$-$electrons atomic or ionic system, is defined as (in atomic units (a.u.)), 
\begin{eqnarray}
\label{H_DC}
H_{DC} \; = \; \sum_{s=1}^{N} \left[ c\bm{\alpha}_s \cdot \bm{p}_s + (\beta_s-1)c^2  + V_s^{nuc} \right ]  + \sum_{s<t}^{N}\frac{1}{r_{st}}  \; .
\end{eqnarray}
Here, $\bm{\alpha}$ and $\beta$ are the 4 x 4 Dirac matrices, $\bm{ p}_{s}$ is the momentum operator of the $s^{th}$ electron, $V^{nuc}$ is the electron-nucleus interaction potential and $c$ is the speed of light. The last term represents the electron-electron Coulomb interaction. \\
The atomic state functions, describing the energy states for $N$ electrons, are given as a linear combination of the configuration state functions (CSFs),

\begin{eqnarray}
\big|\Psi\big> \; = \; \sum_{s=1}^{N_{C S F}}e_{s}\big|\gamma_{s}P J M_J\big> , 
\end{eqnarray}
where e$_s$ are the mixing coeﬃcients. $P$, $J$ and $M_J$ represent the parity, total angular momentum quantum number and its associated magnetic quantum number, respectively. $\gamma$ refers to the occupation number, coupling scheme and all other appropriate quantum labels that are required for the unique specification of a CSF.\\ 
These CSFs are setup from anti-symmetric products of Dirac's one-electron orbitals,

\begin{eqnarray}
\big|\phi\big> \; = \;  \frac{1}{r} \begin{pmatrix}
F_{n\kappa}(r)\chi_{\kappa m}(\Omega)\\
\iota G_{n\kappa}(r)\chi_{-\kappa m}(\Omega)
\end{pmatrix},
\end{eqnarray}
where $F_{n\kappa}$ and $G_{n\kappa}$ are the large and small parts of the radial wavefunction, $n$, and $\kappa$ refer to the principle and relativistic quantum numbers, respectively and $\chi$ denotes the spinor spherical harmonics.\\ 
Employing the variational principle method and the relativistic self consistent field (RSCF) procedure, at first, the radial components of one electron orbitals are optimised to self consistency. Utilizing these orbitals, the mixing coefficients can be determined by diagonalization of the Hamiltonian (equation(\ref{H_DC})).\\
Further, to include the corrections due to transverse photon exchange, the Breit interaction term \cite{mann1971breit}, as given below, is added to the Dirac-Coulomb Hamiltonian,

\begin{eqnarray}
H_{B r e i t}\;=\;-\sum_{s<t}^{N}\left[\bm{\alpha}_{s}\cdot \bm{\alpha}_{t}\frac{\cos({\omega_{s t}  r_{s t}/c})}{r_{s t}}+
(\bm{\alpha}_{s}\cdot\bm{\nabla}_{s})(\bm{\alpha}_{t}\cdot\bm{\nabla}_{t})
\frac{\cos({\omega_{s t}  r_{s t}/c})-1}{\omega_{s t}^{2} r_{s t}/c^{2}}\right],
\end{eqnarray}
where, $\omega_{st}$ denotes the virtual photon frequency. We have also incorporated quantum electrodynamic (QED) effects, viz., self-energy, vacuum polarization, and specific mass shift corrections in equation (\ref{H_DC}) while performing the RCI calculations \cite{mackenzie1980program}. Only the mixing coefficients are computed by diagonalizing the Hamiltonian matrix, and the Dirac orbitals obtained from the previous MCDF procedure are kept unchanged.\\
Once the ASFs are determined, all the relevant electronic structure properties like transition rates, hyperfine structure constants, Landé $g_J$ factors, and isotope shift factors can be calculated as described in the following sections.
\subsection{Radiative transition parameters}
For a transition between two states, the transition parameters, viz., transition rates and oscillator strengths, can be given in terms of the reduced matrix elements of    operator $(O)$ corresponding to the multipole expansion of electron-photon interaction  \cite{grant1974gauge},  
\begin{eqnarray}
\big<u\big|\big|O_\xi^q\big|\big|l\big>,
\end{eqnarray}
where, $\xi$ is rank of the operator $(O)$. $\xi = 0$ or 1, defines the electric or magnetic type of multipole, $q = 1$ and 2 denote dipole and quadrupole transitions, $u$ and $l$ refer to  wavefunctions of the upper and lower states, respectively.
With the use of Racah algebra, these matrices can be further simplified into a weighted sum over radial integrals, and the weights can be acquired from the angular integration \cite{gaigalas2001program}. In the present work, we have considered only the dipole and quadrupole transitions corresponding to $q = 1$ and 2.

\subsection{Hyperfine structures and Landé $g_{J}$ factors}

The interaction of electrons with the electromagnetic multipole moments of nucleus, leads to splitting of the fine-structure levels and is known as the hyperfine effect. The corresponding interaction Hamiltonian in terms of the spherical tensor operators  $\bm{T}^{(k)}$ and $\bm{R}^{(k)}$, in the electron and nucleus spaces, \cite{osti_4571333}, can be expressed as,
\begin{eqnarray}
\label{hhfs}
H_{hfs} \;= \; \sum_{k \geq 1}\bm{T}^{(k)}\bm{\cdot}\bm{R}^{(k)},
\end{eqnarray}
where the rank $k$ = 1 and 2 refer, respectively, to the magnetic dipole and electric
quadrupole interactions and contribute maximum to the hyperfine structures. Hence, the further leading order terms in k can be safely neglected. The $N$-electron tensor operators \bm{$T$}$^{(1)}$ and \bm{$T$}$^{(2)}$ can be written in terms of one-electron tensor operators $\bm{t}^{(k)}$, i.e.,
\begin{eqnarray}
\bm{T}^{(1)} \;
= \; \sum_{s=1}^{N}\bm{t}^{(1)}(s)
= \; -\iota\alpha \sum_{s=1}^N {\bm{\alpha}_{s}\bm{\cdot}\bm{L}_s}X^{(1)}(s)r_s^{-2},
\end{eqnarray}
and
\begin{eqnarray}
\bm{T}^{(2)}\;
= \; \sum_{s=1}^{N}\bm{t}^{(2)}(s)
=\; -\sum_{s=1}^N X^{(2)}(s)r_s^{-3}.
\end{eqnarray}
Here, $\alpha$ is the fine structure constant, $\iota$ refers to the imaginary term, $\bm{L}_s$ denotes the orbital angular momentum operator for the s$^{th}$ electron and $X^{(k)}$ is the spherical tensor whose $m^{th}$ component can be written as,
\begin{eqnarray}
X_m^{(k)} \;  = \; \sqrt{\frac{4\pi}{2k + 1}}Y_{km},
\end{eqnarray}
where $Y_{km}$ are the spherical harmonics.\\
The tensor operators in the nuclear space $(\bm{R}^{(k)})$ are related to the nuclear magnetic dipole moment $\mu_I$ and electric quadrupole moment $Q_I$ as follows \cite{jonsson1996hfs92}, 
\begin{eqnarray}
\mu_I  \; = \; \big<\tau \pi IM_I\big|\bm{R}^{(1)}\big|\tau \pi IM_I\big>,
\end{eqnarray}
and 
\begin{eqnarray}
Q_I \; = \;
\big<\tau \pi I M_I\big|\bm{R}^{(2)}\big|\tau \pi I M_I\big>,
\end{eqnarray}
for $M_I$ = $I$. Here $|\tau \pi IM_I\big>$ denotes the nuclear wave function, $I$ is the nuclear spin quantum number and $M_I$ is the magnetic quantum number corresponding to $I$. $\pi$ represents the parity of the state and $\tau$ includes all the other quantum numbers necessary to uniquely represent a state.\\
Further, the interaction operator in equation (\ref{hhfs}) does not commute with $I$ and $J$ but commutes with the total atomic angular momentum $F$, which is obtained by the coupling of $J$ and $I$, i.e., $F=I+J$.
%
%
%
Now, using the perturbation theory and considering only the diagonal hyperfine interaction terms, the total interaction energy can be expressed as,
\begin{eqnarray}
\big<\tau\gamma \pi P I J F M_F\big| \bm{T}^{(1)}\bm{\cdot}\bm{R}^{(1)} + \bm{T}^{(2)}\bm{\cdot}\bm{R}^{(2)} \big|\tau\gamma \pi P IJFM_F\big>,
\end{eqnarray}
where 
$M_F$ is the magnetic quantum number associated with $F$ and $\big|\tau\gamma \pi P IJFM_F\big>$ is the total wave function for the coupled system.
%
%
%
%
Now employing the Wigner-Eckart theorem, the respective terms for the magnetic dipole $(E_{M1})$ and electric quadrupole $(E_{E2})$ interaction energies  are given by,
\begin{eqnarray}
E_{M1}(J)
\; = \;
(-1)^{I+J+F}\begin{Bmatrix}I &J& F\\
J& I &1
\end{Bmatrix}\sqrt{(2J+1)(2I+1)}\big<\gamma P  J\big|\big|\bm{T}^{(1)}\big|\big|\gamma  P J\big>\big<\tau \pi  I\big|\big|\bm{R}^{(1)}\big|\big|\tau \pi I\big>,
\end{eqnarray}
and 
\begin{eqnarray}
E_{E2}(J)
\; = \;
(-1)^{I+J+F}\begin{Bmatrix}I &J& F\\
J& I &2
\end{Bmatrix}\sqrt{(2J+1)(2I+1)}\big<\gamma P  J\big|\big|\bm{T}^{(2)}\big|\big|\gamma  P J\big>\big<\tau \pi  I\big|\big|\bm{R}^{(2)}\big|\big|\tau \pi I\big>.
\end{eqnarray}
Usually, the $F$ dependence are factorized out, and energies are represented in terms of hyperfine constants A$_J$ and B$_J$ \cite{jonsson1996hfs92},
\begin{eqnarray}
A_J\; = \; \frac{\mu_I}{I}\frac{1}{{\sqrt{(J(J+1))}}}\big<\gamma P J\big|\big|\bm{T}^{(1)}\big|\big|\gamma P J\big>,
\end{eqnarray}
\begin{eqnarray}
B_J\; = \;2Q_I\frac{\sqrt{J(2J-1)}}{\sqrt{(J+1)(2J+3)}}\big< \gamma P J\big|\big|\bm{T}^{(2)}\big|\big|\gamma P J\big>.
\end{eqnarray}
Further, the Landé $g_J$ factor determines the splitting of the magnetic levels in the presence of external magnetic field and can be written as \cite{verdebout2014hyperfine}, 
\begin{eqnarray}
g_J \; = \;
\frac{2}{\sqrt{J(J+1)}} \big< \gamma P J  \big\| \sum_{s=1}^N  \left[ -\iota \frac{\sqrt{2}}{2 \alpha^2} r_s \big( \bm{\alpha}_s\bm{X}_s^{(1)} \big )^{(1)}  + \frac{g_{spin}-2}{2} \beta_s \Gamma_s\right] \big\|\gamma P J\big> \:,
\end{eqnarray}
where $\Gamma$ is the relativistic spin matrix and $g_{spin}$ = 2.00232.
\subsection{Isotope shift}
A specific element may have several isotopes that differ in the number of neutrons inside their respective nuclei. Thus, different isotopes differ in their nuclear masses and volume of the nucleus. These two effects cause a slight difference in the electronic energy states of the isotopes.
Shifting in the energy levels due to nuclear mass is called the mass shift (MS), while that due to nuclear volume is known as the field shift (FS).

For a given atomic state $s$, the FS due to variation in the mean square charge radii $(\big<r^2\big>)$ of the two isotopes having mass numbers $A$ and $A^\prime$,  is given by \cite{blundell1987reformulation, torbohm1985state},
\begin{eqnarray}
\delta E_{A,A^\prime}^s \; = \; F^s
 ( \langle r^2 \rangle _A \;  - \; \langle r^2 \rangle _{A^\prime} ) \; ,
\end{eqnarray}
where 
%
%
$F^s$ is known as the field shift factor and can be written  as,
\begin{eqnarray}
F^s \; = \; \frac{2\pi}{3}\left(\frac{Ze^2}{4\pi\epsilon_o}\right)\Delta|\Psi(\bm{O})|^2_s \; . 
\end{eqnarray}
Here $\Delta|\Psi(\bm{O})|^2_s$ is the electron probability density at origin of the state $s$.\\
The Mass shift of an energy level $i$, of two isotopes with finite nuclear masses $M$ and $M^\prime$ can be written in terms of normal mass shift (NMS) and specific mass shift (SMS), i.e., 
 \begin{eqnarray}
 \delta E_{i,MS}^{A,A'} = \delta E_{i,NMS}^{A,A'} - 
 \delta E_{i,SMS}^{A,A'} = \frac{M-M'}{MM'}K_{MS}\hspace{3mm} ,
 \end{eqnarray}
 where $K_{MS}$ is the electronic factor defined as,
 \begin{eqnarray}
  \label{kms}
\frac{K_{MS}}{M} \; = \; \langle \gamma P J M_J\big|H_{MS}\big|\gamma P J M_J \rangle \hspace{3mm}.
 \end{eqnarray}
$H_{MS}$ in the above equation refers to the mass shift correction operator and given as, \cite{shabaev1994relativistic},
  \begin{eqnarray}
 \label{hms}
 H_{MS} = \frac{1}{2M} \sum_{s,t}^N \left (
 \bm{p_s \cdot p_t} - \frac{\alpha Z}{r_s} 
 \left(\bm{\alpha_{s}}+ 
 \frac{(\bm{\alpha_{s}\cdot r_{s}})\bm{r_s}}{r_s^2}
 \right)\cdot \bm{p_t}
 \right) .
 \end{eqnarray}
$H_{MS}$ can be further split into one body (s = t) term, which corresponds to the NMS, and two body term $(s \neq t)$, which is called the SMS. Finally, we can write the total frequency shift for a particular transition $\eta$ as, 
\begin{eqnarray}
\delta\nu_{A,A^\prime}^\eta \; = \; \hspace{1mm}\delta\nu_{A,A^\prime}^{\eta,NMS} + \hspace{1mm}\delta\nu_{A,A^\prime}^{\eta,SMS} + \hspace{1mm}\delta\nu_{A,A^\prime}^{\eta,FS} \hspace{3mm}.
\end{eqnarray}

\section{Computational Procedure}
In this work, the calculations are performed for the $1s^2$ and $1snl\hspace{1mm}(n=2-8$ and $l=0-7)$ configurations of Sc XX. Based on parity, these states are divided into odd and even parity sets, which defines the multi reference  (MR) set for the even and odd space. There are 63 even states and 64 odd states. All the further calculations for generating the ASFs are done independently for both the parity sets. The first step includes the MCDF calculations for the MR set, which are performed in the extended optimal level (EOL) scheme \cite{dyall1989grasp}. Here, all the orbitals of the MR set are optimized.\\ 
The second step is to expand CSFs of a given parity $P$ and total angular momentum quantum number $J$ using the restricted active set approach. In this technique, the CSFs  are built from the angular coupling of configurations generated by single and double substitutions of electrons from reference set orbitals to the active space (AS) orbitals. In our case we have considered active space up to $n$ = 10 and $l=0-7$. 
To keep a check on convergence, we split AS into layers of the same principal quantum number $n$ and corresponding orbital quantum numbers $l$. We define these AS as, AS$\{9\}  = \{n = 0 - 9, l = s, p, d, f ,g, h, i, k\}$ and AS$\{10\} =  \{n = 0 - 10, l = s, p, d, f ,g, h, i, k\}$. The number of CSFs generated for AS$\{9\}$ are 6104 for even states and 6512 for odd states, while those for AS$\{10\}$ are 8824 for even states and 9472 for odd states. Similar MCDF calculations, as for MR set, are performed for AS \{9\}, where only the newly added orbitals are optimized. Likewise, for AS \{10\}, AS \{9\} orbitals are kept frozen, and only the new orbitals are optimized. After this, the RCI calculations are performed to include the Breit interaction and the leading QED corrections.
Further, to identify the generated states, the transformation from $jj$-coupling to $LSJ$- coupling is performed.
\\
Since the odd and even states are optimised separately, this makes the calculations for transition rates complex due to non-orthogonality of the final and initial states. Therefore, to ease this we performed the biorthogonal transformations of these states.

The relativistic corrected specific mass shift, normal mass shift and field shift are calculated using RIS4. Though RIS4 is a sub program written for GRASP2K$\_$v1$\_$1 \cite{jonsson2013new}, but we did the required modifications so that it can be used in alliance with GRASP2018 code. 

\section{Results and Discussion}
\subsection{Energy levels}
\begin{figure}[H]
\includegraphics[width=10 cm]{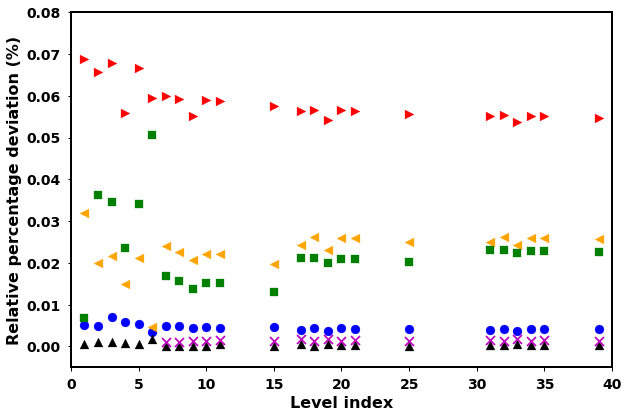}
\caption{With respect to the NIST values, the absolute relative percentage deviation in the energies from the present work (sphere), Si et al \cite{si2016energy} (triangle$\_$up), Massacrier and Artru \cite{massacrier2012extensive} (square), Yerokhin and Surzhykov \cite{yerokhin2019theoretical} (cross), Aggarwal and Keenan \cite{aggarwal2012energy}: MCDF (triangle$\_$right) and FAC-RCI (triangle$\_$left).\label{Fig 1}}
\end{figure}

In Table 1, we listed our calculated energies for all the 127 levels of He-like Sc XX, having configurations $1snl$ with $n=1-8$ and $l=0$ to $n-1$. These results are compared  with the corresponding values from NIST \cite{NIST} and previous works based on RCI \cite{massacrier2012extensive,aggarwal2012energy,yerokhin2019theoretical}, MCDF \cite{aggarwal2012energy} and MBPT \cite{si2016energy} methods. 
For analysis, we calculated the absolute percentage deviation of the present energies and the other's results with respect to the values compiled in the NIST database, where only 24 energy levels are available for comparison. We have shown these percentage deviations in Figure \ref{Fig 1}. It is observed that the absolute mean difference of the present values is 0.004$\%$, while the $1s2p$ $^3P_1$ level shows the maximum deviation of 0.0069 $\%$. Our energies also agree extremely well with those of \cite{si2016energy} and \cite{yerokhin2019theoretical} lying within 0.006$\%$. 
Although, Massacrier and Artru \cite{massacrier2012extensive} have provided energies up to $n=10$ and $l=0$ to $(n-1)$, but their percentage deviation with NIST values is 0.022$\%$, which is more as compared to our calculations. Thus, the above comparison validates the accuracy of our calculated energies. 


\setlength{\tabcolsep}{1pt}
\begin{longtable}{|l|l|l|l|l|l|l|l|}
\caption{Comparison of the energies (in cm$^{-1}$) for the lowest 127 levels of He-like Sc XX .\label{tab1}} \\
\hline \multicolumn{1}{|c|}{\textbf{State }} & \multicolumn{1}{c|}{\textbf{Our work }} & \multicolumn{1}{c|}{\textbf{NIST}}& \multicolumn{1}{c|}{\textbf{FAC-MBPT\cite{si2016energy}}}& \multicolumn{1}{c|}{\textbf{MCDF\cite{aggarwal2012energy}}}& \multicolumn{1}{c|}{\textbf{FAC-RCI\cite{aggarwal2012energy} }}& \multicolumn{1}{c|}{\textbf{RCI\cite{yerokhin2019theoretical} }}& \multicolumn{1}{c|}{\textbf{FAC-RCI\cite{massacrier2012extensive} }}  \\ \hline 
\endfirsthead
\multicolumn{3}{c}%
{{\bfseries \tablename\ \thetable{} -- continued from previous page}} \\
\hline \multicolumn{1}{|c|}{\textbf{State }} & \multicolumn{1}{c|}{\textbf{Our work }} & \multicolumn{1}{c|}{\textbf{NIST}}& \multicolumn{1}{c|}{\textbf{FAC-MBPT\cite{si2016energy}}}& \multicolumn{1}{c|}{\textbf{MCDF\cite{aggarwal2012energy}}}& \multicolumn{1}{c|}{\textbf{FAC-RCI\cite{aggarwal2012energy} }}& \multicolumn{1}{c|}{\textbf{RCI\cite{yerokhin2019theoretical} }}& \multicolumn{1}{c|}{\textbf{FAC-RCI\cite{massacrier2012extensive} }} \\ \hline 
\endhead
\hline \multicolumn{3}{|r|}{{Continued on next page}} \\ \hline
\endfoot
\hline \hline
\endlastfoot
$1s^2 ~^1S_{   0}$ &        0 &        0 &        0 &               0 &             0 &        0 &        0\\
$1s\,2s~^3S_{   1}$ & 34446345 & 34448120 & 34448308 & 34424402 & 34437132 &   & 34450491 \\
$1s\,2p~^3P_{   0}$ & 34627080 & 34628770 & 34629173 & 34606014 &      34621818 &          & 34641298 \\
$1s\,2p~^3P_{   1}$ & 34636133 & 34638550 & 34638223 &  34615063 &      34631061 &          & 34650491 \\
$1s\,2s~^1S_{   0}$ & 34643357 & 34645360 & 34645074 &  34626020 &      34640186 &          & 34653551 \\
$1s\,2p~^3P_{   2}$ & 34680911 & 34682810 & 34683009 & 34659727 &      34675434 &          & 34694642 \\
$1s\,2p~^1P_{   1}$ & 34803824 & 34805000 & 34805659 &        34784308 &      34803372 &          & 34822627 \\
$1s\,3s~^3S_{   1}$ & 40788626 & 40790620 & 40790661 &        40766206 &      40780824 & 40791080 & 40783716 \\
$1s\,3p~^3P_{   0}$ & 40838511 & 40840530 & 40840500 &        40816289 &      40831283 & 40840930 & 40834109 \\
$1s\,3s~^1S_{   0}$ & 40840663 & 40842480 & 40842462 &        40819913 &      40834018 & 40843010 & 40836910 \\
$1s\,3p~^3P_{   1}$ & 40841191 & 40843100 & 40843177 &        40818982 &      40834033 & 40843609 & 40836850 \\
$1s\,3p~^3P_{   2}$ & 40854547 & 40856320 & 40856540 &        40832298 &      40847324 & 40856964 & 40850113 \\
$1s\,3d~^3D_{   2}$ & 40881891 &          & 40883855 &        40859407 &      40873552 &          & 40879646 \\
$1s\,3d~^3D_{   1}$ & 40882043 &          & 40884013 &        40859538 &      40873640 &          & 40879735 \\
$1s\,3d~^3D_{   3}$ & 40887189 &          & 40889156 &        40864665 &      40878754 &          & 40884782 \\
$1s\,3p~^1P_{   1}$ & 40887768 & 40889690 & 40889644 &        40866199 &      40881576 & 40890176 & 40884369 \\
$1s\,3d~^1D_{   2}$ & 40888749 &          & 40890702 &        40866289 &      40880481 &          & 40886511 \\
$1s\,4s~^3S_{   1}$ & 42981651 & 42983370 & 42983634 &        42959155 &      42972908 & 42984110 & 42974305 \\
$1s\,4p~^3P_{   0}$ & 43002186 & 43004100 & 43004149 &        42979744 &      42992825 & 43004621 & 42995000 \\
$1s\,4s~^1S_{   0}$ & 43002768 & 43004390 & 43004644 &        42981097 &      42994436 & 43005152 & 42995827 \\
$1s\,4p~^3P_{   1}$ & 43003315 & 43005180 & 43005276 &        42980885 &      42993987 & 43005749 & 42996156 \\
$1s\,4p~^3P_{   2}$ & 43008958 & 43010770 & 43010923 &        42986508 &      42999626 & 43011392 & 43001761 \\
$1s\,4d~^3D_{   2}$ & 43020237 &          & 43022192 &        42997694 &      43012127 &          & 43014294 \\
$1s\,4d~^3D_{   1}$ & 43020246 &          & 43022202 &        42997690 &      43012104 &          & 43014273 \\
$1s\,4d~^3D_{   3}$ & 43022416 &          & 43024372 &        42999854 &      43014241 &          & 43016404 \\
$1s\,4p~^1P_{   1}$ & 43022539 & 43024380 & 43024449 &        43000447 &      43013585 & 43024959 & 43015723 \\
$1s\,4f~^3F_{   3}$ & 43023120 &          & 43025071 &        43000524 &      43013939 &          & 43015970 \\
$1s\,4d~^1D_{   2}$ & 43023216 &          & 43025169 &        43000692 &      43015135 &          & 43017295 \\
$1s\,4f~^3F_{   2}$ & 43023257 &          & 43025210 &        43000661 &      43014076 &          & 43016107 \\
$1s\,4f~^3F_{   4}$ & 43024350 &          & 43026302 &        43001750 &      43015162 &          & 43017189 \\
$1s\,4f~^1F_{   3}$ & 43024462 &          & 43026413 &        43001864 &      43015279 &          & 43017303 \\
$1s\,5s~^3S_{   1}$ & 43990475 & 43992240 & 43992441 &        43967946 &      43981267 & 43992929 & 43982130 \\
$1s\,5p~^3P_{   0}$ & 44000864 & 44002720 & 44002819 &        43978354 &      43991220 & 44003303 & 43992599 \\
$1s\,5s~^1S_{   0}$ & 44001105 & 44002740 & 44002997 &        43979094 &      43992077 & 44003495 & 43992936 \\
$1s\,5p~^3P_{   1}$ & 44001442 & 44003270 & 44003396 &        43978940 &      43991821 & 44003880 & 43993191 \\
$1s\,5p~^3P_{   2}$ & 44004331 & 44006140 & 44006287 &        43981820 &      43994703 & 44006769 & 43996061 \\
$1s\,5d~^3D_{   1}$ & 44010037 &          & 44011990 &        43987466 &      44001673 &          & 44002442 \\
$1s\,5d~^3D_{   2}$ & 44010044 &          & 44011997 &        43987479 &      44001696 &          & 44002465 \\
$1s\,5d~^3D_{   3}$ & 44011148 &          & 44013101 &        43988575 &      44002765 &          & 44003535 \\
$1s\,5p~^1P_{   1}$ & 44011190 & 44013010 & 44013109 &        43988910 &      44001740 & 44013613 & 44003097 \\
$1s\,5f~^3F_{   3}$ & 44011540 &          & 44013491 &        43988950 &      44002322 &          & 44003355 \\
$1s\,5d~^1D_{   2}$ & 44011589 &          & 44013540 &        43989036 &      44003256 &          & 44004027 \\
$1s\,5f~^3F_{   2}$ & 44011608 &          & 44013561 &        43989017 &      44002393 &          & 44003423 \\
$1s\,5f~^3F_{   4}$ & 44012168 &          & 44014120 &        43989576 &      44002949 &          & 44003977 \\
$1s\,5g~^3G_{   4}$ & 44012185 &          & 44014136 &        43989590 &      44002931 &          & 44003938 \\
$1s\,5g~^3G_{   3}$ & 44012227 &          & 44014178 &        43989630 &      44002975 &          & 44003980 \\
$1s\,5f~^1F_{   3}$ & 44012228 &          & 44014179 &        43989637 &      44003012 &          & 44004038 \\
$1s\,5g~^3G_{   5}$ & 44012563 &          & 44014514 &        43989965 &      44003310 &          & 44004314 \\
$1s\,5g~^1G_{   4}$ & 44012596 &          & 44014547 &        43989998 &      44003347 &          & 44004348 \\
$1s\,6s~^3S_{   1}$ & 44536379 &          & 44538374 &                 &               & 44538830 & 44527483 \\
$1s\,6p~^3P_{   0}$ & 44542351 &          & 44544335 &                 &               & 44544792 & 44533284 \\
$1s\,6s~^1S_{   0}$ & 44542484 &          & 44544420 &                 &               & 44544876 & 44533501 \\
$1s\,6p~^3P_{   1}$ & 44542685 &          & 44544668 &                 &               & 44545125 & 44533840 \\
$1s\,6p~^3P_{   2}$ & 44544356 &          & 44546342 &                 &               & 44546798 & 44535565 \\
$1s\,6d~^3D_{   1}$ & 44547634 &          & 44549609 &                 &               &          & 44539193 \\
$1s\,6d~^3D_{   2}$ & 44547642 &          & 44549616 &                 &               &          & 44539297 \\
$1s\,6d~^3D_{   3}$ & 44548277 &          & 44550252 &                 &               &          & 44539911 \\
$1s\,6p~^1P_{   1}$ & 44548301 &          & 44550258 &                 &               & 44550724 & 44539595 \\
$1s\,6f~^3F_{   3}$ & 44548514 &          & 44550476 &                 &               &          & 44539819 \\
$1s\,6d~^1D_{   2}$ & 44548542 &          & 44550515 &                 &               &          & 44540155 \\
$1s\,6f~^3F_{   2}$ & 44548553 &          & 44550515 &                 &               &          & 44539815 \\
$1s\,6f~^3F_{   4}$ & 44548877 &          & 44550839 &                 &               &          & 44540175 \\
$1s\,6g~^3G_{   4}$ & 44548887 &          & 44550842 &                 &               &          & 44540143 \\
$1s\,6g~^3G_{   3}$ & 44548912 &          & 44550866 &                 &               &          & 44540143 \\
$1s\,6f~^1F_{   3}$ & 44548912 &          & 44550874 &                 &               &          & 44540179 \\
$1s\,6g~^3G_{   5}$ & 44549106 &          & 44551060 &                 &               &          & 44540358 \\
$1s\,6h~^3H_{   5}$ & 44549109 &          & 44551061 &                 &               &          & 44540351 \\
$1s\,6h~^3H_{   4}$ & 44549124 &          & 44551076 &                 &               &          & 44540351 \\
$1s\,6g~^1G_{   4}$ & 44549126 &          & 44551080 &                 &               &          & 44540359 \\
$1s\,6h~^3H_{   6}$ & 44549254 &          & 44551206 &                 &               &          & 44540495 \\
$1s\,6h~^1H_{   5}$ & 44549268 &          & 44551219 &                 &               &          & 44540495 \\
$1s\,7s~^3S_{   1}$ & 44864673 &          &          &                 &               & 44867123 & 44855576 \\
$1s\,7p~^3P_{   0}$ & 44868418 &          &          &                 &               & 44870860 & 44859210 \\
$1s\,7s~^1S_{   0}$ & 44868509 &          &          &                 &               & 44870901 & 44859338 \\
$1s\,7p~^3P_{   1}$ & 44868629 &          &          &                 &               & 44871070 & 44859559 \\
$1s\,7p~^3P_{   2}$ & 44869681 &          &          &                 &               & 44872123 & 44860644 \\
$1s\,7d~^3D_{   1}$ & 44871736 &          &          &                 &               &          & 44862918 \\
$1s\,7d~^3D_{   2}$ & 44871742 &          &          &                 &               &          & 44862984 \\
$1s\,7d~^3D_{   3}$ & 44872140 &          &          &                 &               &          & 44863370 \\
$1s\,7p~^1P_{   1}$ & 44872160 &          &          &                 &               & 44874584 & 44863168 \\
$1s\,7f~^3F_{   3}$ & 44872293 &          &          &                 &               &          & 44863319 \\
$1s\,7d~^1D_{   2}$ & 44872311 &          &          &                 &               &          & 44863529 \\
$1s\,7f~^3F_{   2}$ & 44872317 &          &          &                 &               &          & 44863317 \\
$1s\,7f~^3F_{   4}$ & 44872521 &          &          &                 &               &          & 44863543 \\
$1s\,7g~^3G_{   4}$ & 44872528 &          &          &                 &               &          & 44863521 \\
$1s\,7g~^3G_{   3}$ & 44872544 &          &          &                 &               &          & 44863521 \\
$1s\,7f~^1F_{   3}$ & 44872544 &          &          &                 &               &          & 44863547 \\
$1s\,7g~^3G_{   5}$ & 44872666 &          &          &                 &               &          & 44863657 \\
$1s\,7h~^3H_{   5}$ & 44872668 &          &          &                 &               &          & 44863652 \\
$1s\,7h~^3H_{   4}$ & 44872678 &          &          &                 &               &          & 44863652 \\
$1s\,7g~^1G_{   4}$ & 44872678 &          &          &                 &               &          & 44863657 \\
$1s\,7h~^3H_{   6}$ & 44872760 &          &          &                 &               &          & 44863742 \\
$1s\,7i~^3I_{   6}$ & 44872760 &          &          &                 &               &          & 44863729 \\
$1s\,7i~^3I_{   5}$ & 44872767 &          &          &                 &               &          & 44863729 \\
$1s\,7h~^1H_{   5}$ & 44872768 &          &          &                 &               &          & 44863742 \\
$1s\,7i~^3I_{   7}$ & 44872825 &          &          &                 &               &          & 44863794 \\
$1s\,7i~^1I_{   6}$ & 44872831 &          &          &                 &               &          & 44863794 \\
$1s\,8s~^3S_{   1}$ & 45077315 &          &          &                 &               &          & 45068146 \\
$1s\,8p~^3P_{   0}$ & 45079840 &          &          &                 &               &          & 45070571 \\
$1s\,8s~^1S_{   0}$ & 45079895 &          &          &                 &               &          & 45070655 \\
$1s\,8p~^3P_{   1}$ & 45079981 &          &          &                 &               &          & 45070804 \\
$1s\,8p~^3P_{   2}$ & 45080685 &          &          &                 &               &          & 45071530 \\
$1s\,8d~^3D_{   1}$ & 45082057 &          &          &                 &               &          & 45073046 \\
$1s\,8d~^3D_{   2}$ & 45082061 &          &          &                 &               &          & 45073091 \\
$1s\,8d~^3D_{   3}$ & 45082327 &          &          &                 &               &          & 45073350 \\
$1s\,8p~^1P_{   1}$ & 45082351 &          &          &                 &               &          & 45073216 \\
$1s\,8f~^3F_{   3}$ & 45082431 &          &          &                 &               &          & 45073318 \\
$1s\,8d~^1D_{   2}$ & 45082443 &          &          &                 &               &          & 45073458 \\
$1s\,8f~^3F_{   2}$ & 45082448 &          &          &                 &               &          & 45073316 \\
$1s\,8f~^3F_{   4}$ & 45082584 &          &          &                 &               &          & 45073468 \\
$1s\,8g~^3G_{   4}$ & 45082589 &          &          &                 &               &          & 45073452 \\
$1s\,8f~^1F_{   3}$ & 45082599 &          &          &                 &               &          & 45073471 \\
$1s\,8g~^3G_{   3}$ & 45082599 &          &          &                 &               &          & 45073452 \\
$1s\,8g~^3G_{   5}$ & 45082681 &          &          &                 &               &          & 45073543 \\
$1s\,8h~^3H_{   5}$ & 45082683 &          &          &                 &               &          & 45073540 \\
$1s\,8h~^3H_{   4}$ & 45082689 &          &          &                 &               &          & 45073540 \\
$1s\,8g~^1G_{   4}$ & 45082689 &          &          &                 &               &          & 45073543 \\
$1s\,8h~^3H_{   6}$ & 45082744 &          &          &                 &               &          & 45073601 \\
$1s\,8i~^3I_{   6}$ & 45082744 &          &          &                 &               &          & 45073591 \\
$1s\,8i~^3I_{   5}$ & 45082749 &          &          &                 &               &          & 45073591 \\
$1s\,8h~^1H_{   5}$ & 45082749 &          &          &                 &               &          & 45073601 \\
$1s\,8i~^3I_{   7}$ & 45082788 &          &          &                 &               &          & 45073635 \\
$1s\,8k~^3K_{   7}$ & 45082789 &          &          &                 &               &          & 45073620 \\
$1s\,8i~^1I_{   6}$ & 45082792 &          &          &                 &               &          & 45073635 \\
$1s\,8k~^3K_{   6}$ & 45082792 &          &          &                 &               &          & 45073620 \\
$1s\,8k~^3K_{   8}$ & 45082821 &          &          &                 &               &          & 45073652 \\
$1s\,8k~^1K_{   7}$ & 45082824 &          &          &                 &               &          & 45073652 \\
\end{longtable}

\subsection{Radiative transition parameters}
Using the MCDF method along with the Breit and QED corrections, we have calculated the transition rates, oscillator strengths and line strengths (S) for all the possible transitions among the 127 levels of He-like Sc XX ion. These calculations have been done in both the gauges, viz., Coulomb and Babushkin gauges, for electric and magnetic dipole as well as quadrupole transitions. Depending on the type of a transition between lower state $l$ and upper state $u$, these parameters are related to each other by the following relations:
\begin{eqnarray}
f_{lu}^{E1}= \frac{303.75 \times S(E1)}{\lambda_{ul}g_l} \hspace{11mm}
\text{and} \hspace{11mm}
A_{lu}^{E1}= \frac{2.0261\times 10^{18} \times S(E1)}{\lambda_{ul}^3g_u},
\end{eqnarray}
%
\begin{eqnarray}
f_{lu}^{E2}= \frac{167.89 \times S(E2)}{\lambda_{ul}^3g_l}\hspace{11mm}
\text{and} \hspace{10mm}
A_{lu}^{E2}= \frac{1.1199 \times10^{18} \times S(E2)}{\lambda_{ul}^5g_u} ,
\end{eqnarray}
%
\begin{eqnarray}
f_{lu}^{M1}= \frac{4.044\times10^{-3} \times S(M1)}{\lambda_{ul}g_l}\hspace{6mm}
\text{and} \hspace{10mm}
A_{lu}^{M1}= \frac{2.6974\times10^{13} \times S(M1)}{\lambda_{ul}^3g_u} ,
\end{eqnarray}
%
\begin{eqnarray}
f_{lu}^{M2}= \frac{2.236\times10^{-3} \times S(M2)} {\lambda_{ul}^3g_l} \hspace{6mm}
\text{and} \hspace{10mm}
A_{lu}^{M2}= \frac{1.4910\times10^{13} \times S(M2)}{\lambda_{ul}^5g_u} .
\end{eqnarray}
\\
Here $\lambda_{ul}$ is the wavelength in \AA , $g_l$ and $g_u$ are the statistical weights of the lower and upper states.  To appraise the uncertainty in the obtained transition rates, we have calculated the dT parameter which is given as, 
\begin{eqnarray}
\label{dt}
 dT = \frac{|A_B - A_C|}{ \text{max} (A_B,A_C) }
\end{eqnarray}
where $A_B$ and $A_C$ are the transition rates in the Babushkin and Coulomb gauges.
\begin{figure}[H]
\includegraphics[width=8cm,height=6cm]{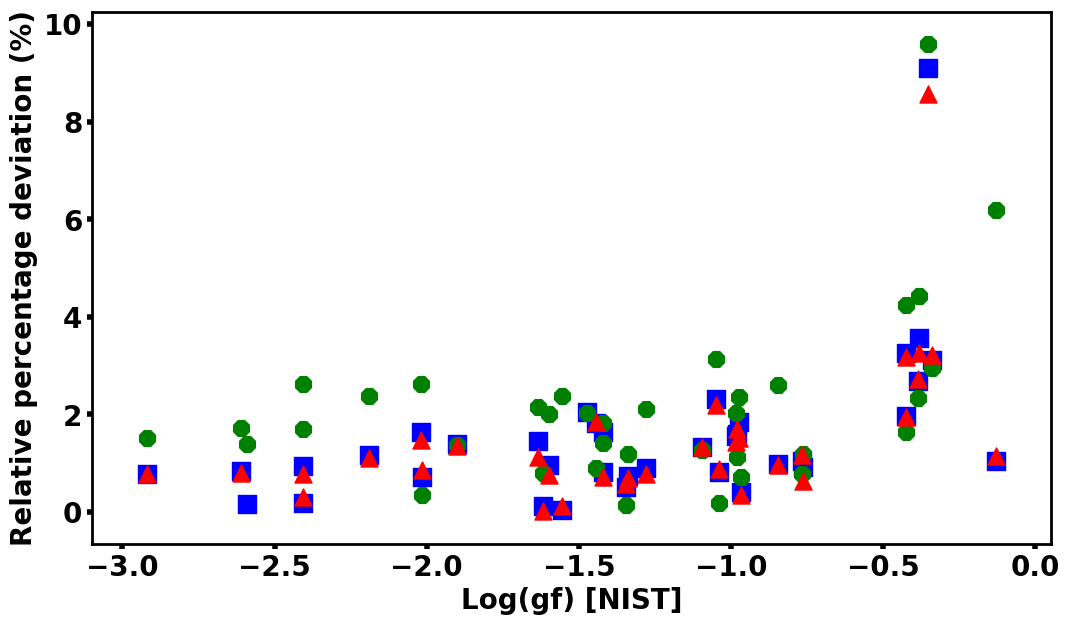}
\caption{Absolute relative percentage difference in weighted oscillator strengths $(\log(gf))$ from the present work (sphere), Si et al. \cite{si2016energy} (triangle$\_$up) and Massacrier and Artru \cite{massacrier2012extensive} (square) with respect to the NIST database values.\label{fig 2}}
\end{figure}
In the present work, the number of E1, E2, M1 and M2 transitions are, respectively, 1635, 1992, 1088 and 2120. However, we do not report here those  E2, M1 and M2 transitions for which the order of the oscillator strength is less than $10^{-10}$. After limiting the order of the f values, the number of lines reduces to 1600, 421 and 530 for E2, M1 and M2 transitions, respectively. 
Table 2 shows the detailed comparison of the present $A$ and $f$ values for E1 transitions with the other calculations \cite{massacrier2012extensive, si2016energy} and the NIST \cite{NIST} line parameters available only for 39 transitions. Out of these 39 NIST values, 37 belong to the E1 transitions while the rest two correspond to M1 and M2 transitions.  
Massacrier and Artru \cite{massacrier2012extensive}, performed the calculations only for E1 transitions and computed the weighted oscillator strengths (log(gf)) for 1214 transitions among the 127 levels considered in this work. Si et al \cite{si2016energy} reported the transition rates, oscillator strengths  and line strengths for a total of 296 transitions among the $1s^2$ to $1snl$ levels for $n=1-6$ and $l=0-(n-1)$.
%
Although the detailed results can be accessed from Table 2, to understand the comparison at a glance, we have presented in Figure \ref{fig 2},  the absolute percentage deviation of all the three theoretical results for $\log(gf)$  with respect to the NIST values.
The mean error in our results is found to be 1.53\% whereas, in the calculations of Si et al  \cite{si2016energy} and Massacrier and Artru \cite{massacrier2012extensive} its values are 1.50\% and  2.20\%, respectively. We observed that the maximum difference of 9.11\% occurs for $1s5p~ ^1P_1$ - $1s4s~^1S_0$ transition. However, all the three theoretical values are within 1\% for this transition. We also found that the uncertainty parameter dT is 0.003 in this case, which further brace the accuracy of our calculations.
\begin{figure}[H]
\captionsetup{justification=centering}
     \begin{tabular}{@{}c@{}}
        \includegraphics[width=6cm,height=5.93cm]{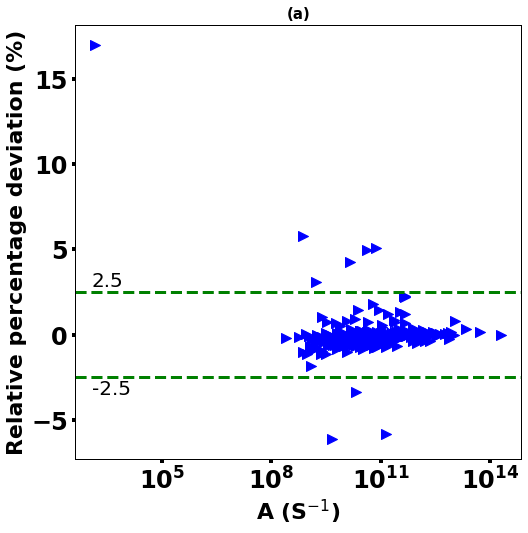}
    \end{tabular}
    \hspace{1em}   
    \begin{tabular}{@{}c@{}}
        \includegraphics[width=6cm,height=5.93cm]{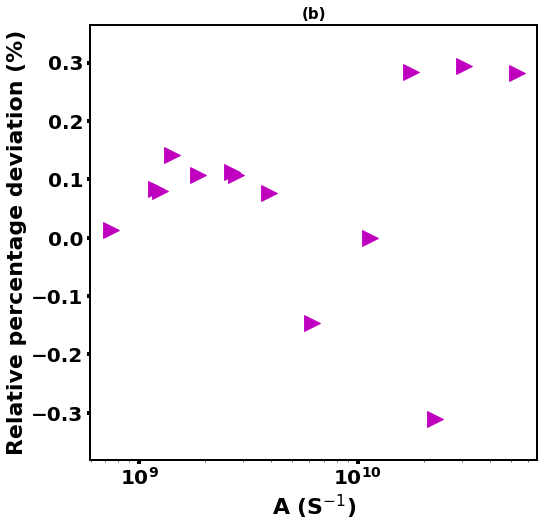}
    \end{tabular}
    \vspace{1pt}
        \caption{Relative percentage difference of our calculated transition rates A (s$^{-1}$) with respect to the results of Si et al.\cite{si2016energy} for (a) E1  and (b) E2 transitions. }
        \label{fig:3}
\end{figure}[H]
Next, we compared our calculated transition rates with the corresponding results for 279 E1 lines reported by \cite{si2016energy} in Figure \ref{fig:3} (a). The two results are in excellent agreement with an absolute mean deviation of 0.5\%. The only exception is $1s2p~^3P_1 - 1s2s~^1S_0$ transition for which the difference is ~17\% in the transition rate and ~5\% in the oscillator strength. The discrepancy could be due to a small value $\left( 1.7 \times 10^{-5} \right )$ of the oscillator strength and a nearly 5\% difference in the wavelengths from the two calculations. The wavelength computed by us is 13842.40 \AA, whereas Si et al. \cite{si2016energy} reported it to be 14597.80 \AA.
Table 3 represents a comparison of our results for E2 transitions with the MBPT calculations of Si et al. \cite{si2016energy} which have been reported only for 14 lines. Both the theoretical results are in excellent agreement. The maximum variation between the transition rates is 0.31\%, with an absolute average difference of 0.15\%, as is easily visible through the comparison displayed graphically in Figure \ref{fig:3} (b). 
%
%
%
Table 4 contains M1 transition parameters. The previous results are available only for two lines. For $1s2s~^3S_1$ to $1s^2~^1S_0$ transition, our calculated transition rate is within 15.66\% and 12.64\% in comparison to the values from the NIST database and Si et al \cite{si2016energy}, respectively. The $f$ value for  $1s2s~^1S_0$ to $1s2s~^3S_1$ transition shows a variation of 16.84\% from the only reported result of \cite{si2016energy}. This relatively large difference observed here in comparison to the results for E1 and E2 transitions (Figures \ref{fig 2} and \ref{fig:3} ) is primarily due to small values of $f$ for the above two transitions. 
\begin{figure}[H]
    \captionsetup{justification=centering}
   \begin{tabular}{@{}c@{}}
        \includegraphics[width=6cm,height=5.93cm]{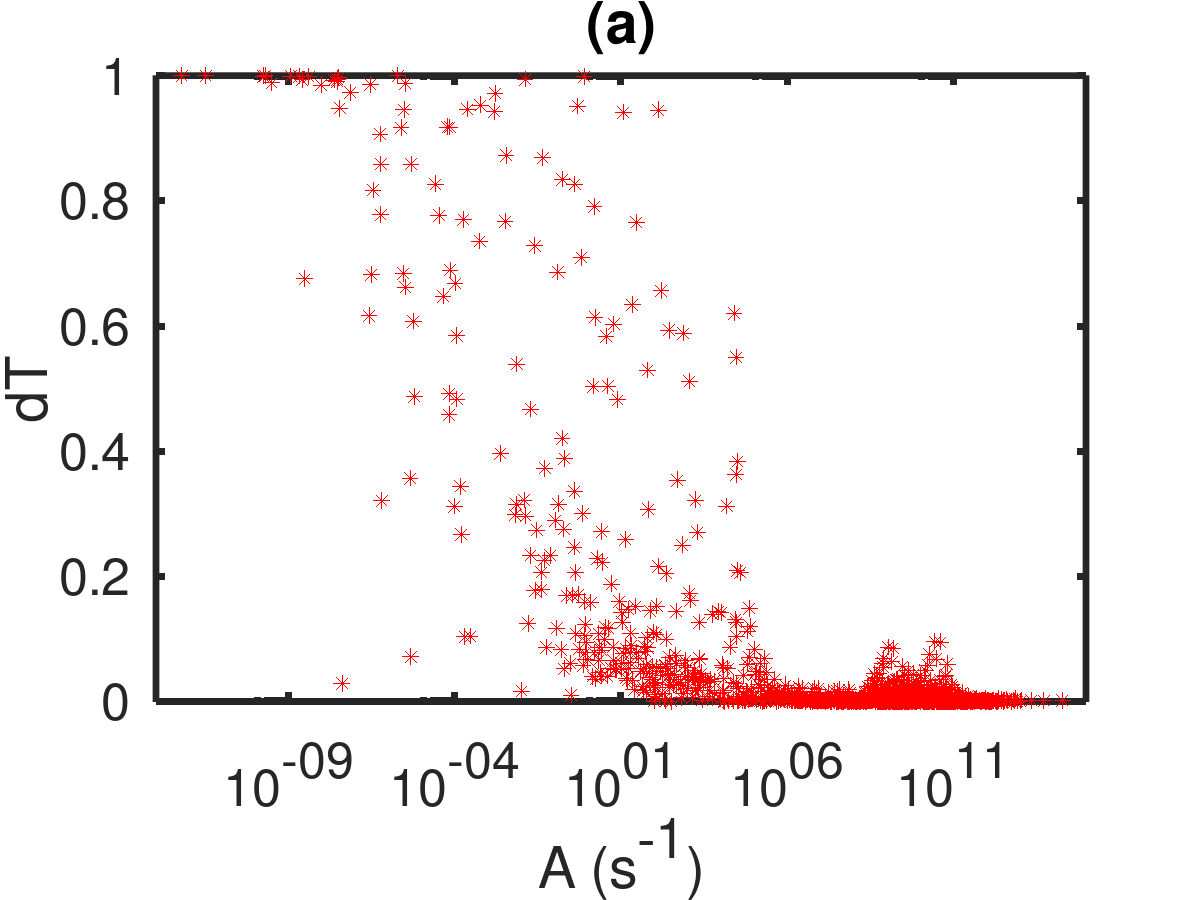}
    \end{tabular}
    \vspace{2pt}    
   \begin{tabular}{@{}c@{}}
        \includegraphics[width=6cm,height=5.93cm]{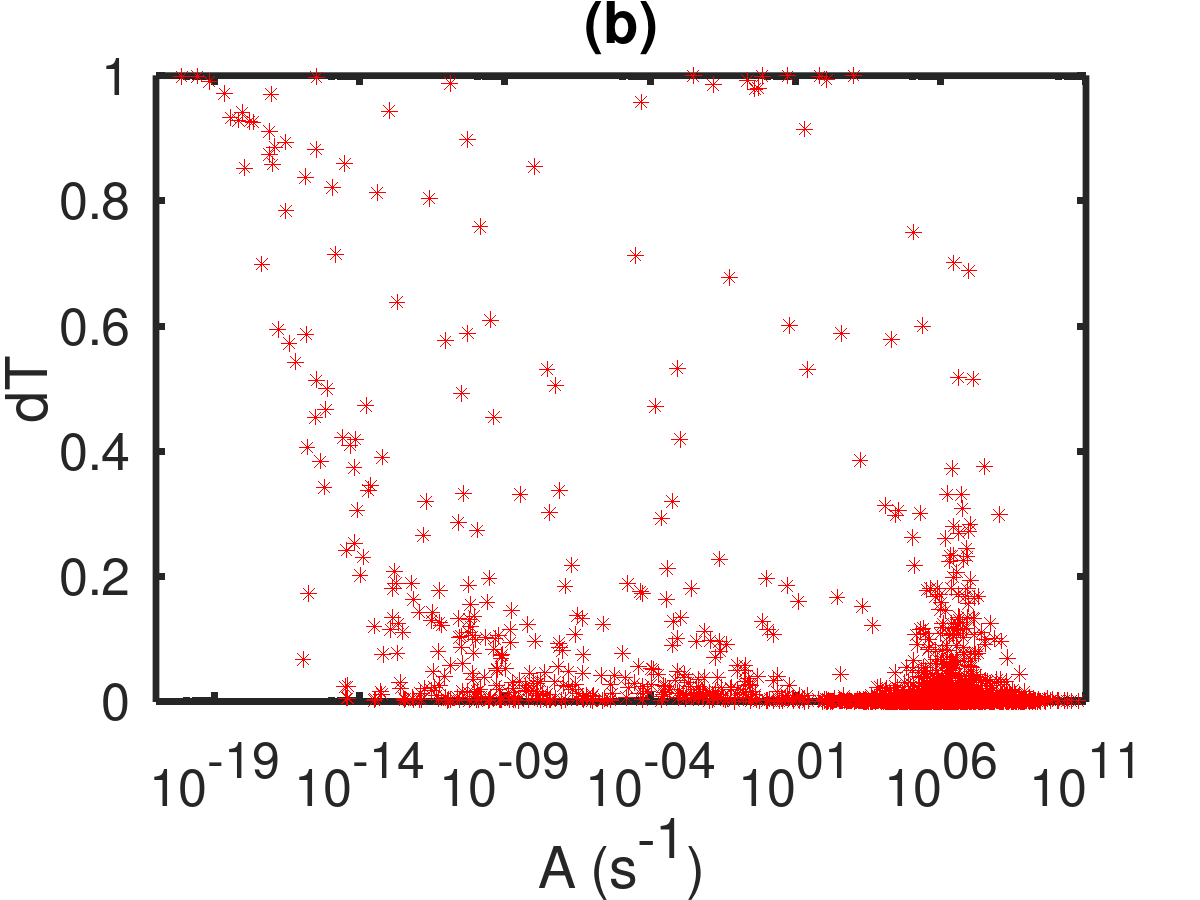}
  \end{tabular}
  \vspace{1pt}
  \caption{dT parameters versus transition rates A (s$^{-1}$) for (a) E1 and (b) E2 transitions.}
        \label{fig:4}
\end{figure}[H]
In Table 5, the results for M2 transitions are tabulated and compared with the corresponding values reported for only one transition between $1s^2~^1S_0$ and $1s2p~^3P_2$ states by the NIST database and Si et al. \cite{si2016energy}. We find an excellent agreement with the other two results. 
To further inspect the accuracy of the present calculations, we determined the dT parameters (equation(\ref{dt})) for E1 and E2 transitions and plotted them with respect to the transition rates in Figures \ref{fig:4} (a) and (b). Only for a handful of transitions, the uncertainty parameter approaches to unity. Majority of the cases have dT below 0.2, while its mean value is 0.05$\%$ for both types of transitions. Thus, a good agreement between the transition rates from the length and velocity gauges further establishes the accuracy of our calculations.
%
\subsection{Lifetimes}
\begin{figure}[H]
\includegraphics[width=9 cm]{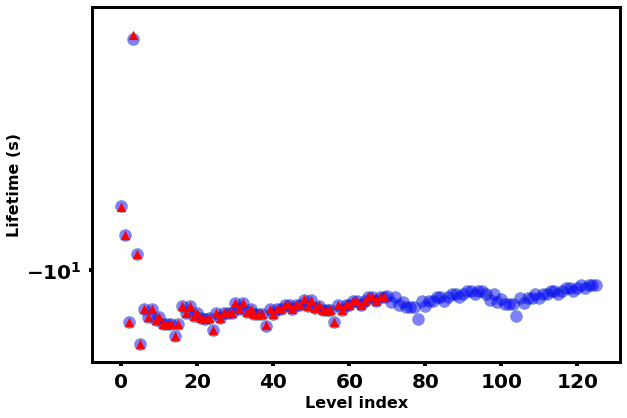}
\caption{Lifetime of the states from the present work (sphere) and Si et al.\cite{si2016energy} (triangle$\_$up).\label{fig 5}} 
\end{figure}

For a level, $l$, the lifetime $\tau$ is defined as the inverse of the spontaneous emission rates of all the possible transitions from any upper level $u$ to that level $l$, and it can be written as,
\begin{eqnarray}
\tau_l \hspace{2mm}  = \hspace{2mm} \frac{1}{ \Sigma_u A_{ul} } . 
\end{eqnarray}
We have calculated the lifetimes of the 127 levels of Sc XX by incorporating the E1, E2, M1, and M2 transitions and listed them in Table 6. The present results are compared with the lifetimes reported only for the lowest 70 levels $(n\leq6; l\leq(n-1))$ by Si et al \cite{si2016energy}. The two theoretical results are in excellent agreement, as can be seen from Figure \ref{fig 5}. 
We report the lifetimes for the higher excited levels 1s7l $(l\leq6)$ and $1s8l$\hspace{1mm} $(l\leq7)$ for the first time, and no previous theoretical or experimental results were available for comparison. We believe that our extensive lifetime calculations will be helpful in future studies.
\subsection{Landé g$_J$-factors and hyperﬁne interaction constants}
We have further used GRASP2018 \cite{fischer2019grasp2018} to calculate the hyperfine constants $A_J$ and $B_J$ and Landé g$_J$-factors. To obtain these values, we have taken the nuclear dipole and quadrupole moments from the compiled values in \cite{stone2005table}.  
In Table 7 the present results for $A_J$, $B_J$ and g$_J$ factors are compared with the MCDF results of Singh et al. \cite{singh2020revised} reported for $1s^2$, $1s2s$, $1s2p$ states only. There is an excellent agreement between the two results for the hyperfine constants. To the best of our knowledge, no theoretical or experimental values are available to compare Landé $g_J$-factors. Since information on the hyperfine structures plays an important role in scandium spectra analysis, our HFS and $g_J$ results will benefit spectra analysis and abundance calculations in astrophysics.
%
%
%
\subsection{Isotope shifts}
We have performed the calculations for the normal mass shift, specific mass shift, and field shift factors using the RIS4 and tabulated these results in atomic units in Table 8. The SMS values of only ground state were calculated by Bhatti et al. \cite{bhatti2001mcdf} in cm$^{-1}$. Hence, to compare with their results, we converted our calculated SMS from a.u. to cm$^{-1}$ using the conversion formula mentioned in \cite{accad1971s}. The present SMS is 7.467 cm$^{-1}$ whereas, Bhatti et al.\cite{bhatti2001mcdf} reported 7.395 cm$^{-1}$ and 7.187 cm$^{-1}$ using the MCDF and multi-configuration Hartree–Fock (MCHF) methods, respectively. Thus, an excellent match can be seen in the specific mass shifts. One reason for the slight difference in the above values could be the conversion in units. We could not find any other experimental or theoretical work to further compare with our results.
\section{Conclusion}
In this study, the energy levels, transition parameters, HFS constants, Landé g$_{J}$ factors, and isotope shifts for He - like Sc XX have been calculated using GRASP2018 and RIS4. These calculations include the lowest 127 levels. We incorporated the relativistic corrections to enhance the quality of the wavefunctions employed in the present work. Most of the forbidden transition parameters, HFSs, Landé g$_J$ factors and ISs for high excited states are reported for the first time, in addition to lifetimes for some of the states. An excellent agreement is obtained on comparing the present results with the NIST values and other theoretical studies, wherever available. The maximum deviation in our calculated energies is only 0.007 \% as compared to other results. Except for a few weak lines, transition parameters for all the allowed and forbidden transitions show a minimal difference from the previous results. Hence, we conclude that our results are comprehensive, accurate, and reliable. We hope that our calculations will help meet the accentuated demand of complete results for Sc XX and will also be useful in the interpretation of observed stellar spectra, high-temperature laboratory plasma, abundance anomalies in Am-Fm stars, galaxies and other astrophysical objects.

\section{Supplementary Materials}
Tables 2 - 8 are provided in the supplementary data.

\section{Author Contributions}
Both the authors have contributed equally.

\section{Funding}
This research has received no external funding.

\section{Acknowledgments}
Shikha Rathi gratefully acknowledges the Ministry of Education (MoE), India, for the research fellowship.

\section{Conflicts of Interest}
The authors declare no conflict of interest.

\section{References}

\bibliography{main.bib}

\begin{thebibliography}{10}
\expandafter\ifx\csname url\endcsname\relax
  \def\url#1{\texttt{#1}}\fi
\expandafter\ifx\csname urlprefix\endcsname\relax\def\urlprefix{URL }\fi
\expandafter\ifx\csname href\endcsname\relax
  \def\href#1#2{#2} \def\path#1{#1}\fi

\bibitem{wahlgren2005inputting}
G.~Wahlgren, Inputting hyperfine structure into synthetic spectrum codes,
  Memorie della Societa Astronomica Italiana Supplementi 8 (2005) 108.

\bibitem{lawler2019transition}
J.~Lawler, C.~Sneden, G.~Nave, M.~Wood, J.~Cowan, et~al., Transition
  probabilities of sc i and sc ii and scandium abundances in the sun, arcturus,
  and hd 84937, The Astrophysical Journal Supplement Series 241~(2) (2019) 21.

\bibitem{prieto2020chemical}
C.~A. Prieto, Chemical composition of the solar surface, Journal of
  Astrophysics and Astronomy 41~(1) (2020) 1--6.

\bibitem{lodders2003solar}
K.~Lodders, Solar system abundances and condensation temperatures of the
  elements, The Astrophysical Journal 591~(2) (2003) 1220.

\bibitem{nissen2016high}
P.~Nissen, High-precision abundances of sc, mn, cu, and ba in solar
  twins-trends of element ratios with stellar age, Astronomy \& Astrophysics
  593 (2016) A65.

\bibitem{nissen1999sc}
P.~Nissen, Y.~Chen, W.~Schuster, G.~Zhao, Sc and mn abundances in disk and
  metal-rich halo stars, arXiv preprint astro-ph/9912269.

\bibitem{sneden2016iron}
C.~Sneden, J.~J. Cowan, C.~Kobayashi, M.~Pignatari, J.~E. Lawler, E.~A.
  Den~Hartog, M.~P. Wood, Iron-group abundances in the metal-poor main-sequence
  turnoff star hd 84937, The Astrophysical Journal 817~(1) (2016) 53.

\bibitem{cohen2004abundances}
J.~G. Cohen, N.~Christlieb, A.~McWilliam, S.~Shectman, I.~Thompson,
  G.~Wasserburg, I.~Ivans, M.~Dehn, T.~Karlsson, J.~Melendez, Abundances in
  very metal-poor dwarf stars, The Astrophysical Journal 612~(2) (2004) 1107.

\bibitem{weisskopf2000chandra}
M.~C. Weisskopf, H.~D. Tananbaum, L.~P. Van~Speybroeck, S.~L. O'Dell, Chandra
  x-ray observatory (cxo): overview, in: X-Ray Optics, Instruments, and
  Missions III, Vol. 4012, International Society for Optics and Photonics,
  2000, pp. 2--16.

\bibitem{reynolds2008youngest}
S.~P. Reynolds, K.~J. Borkowski, D.~A. Green, U.~Hwang, I.~Harrus, R.~Petre,
  The youngest galactic supernova remnant: G1. 9+ 0.3, The Astrophysical
  Journal Letters 680~(1) (2008) L41.

\bibitem{borkowski2010radioactive}
K.~J. Borkowski, S.~P. Reynolds, D.~A. Green, U.~Hwang, R.~Petre,
  K.~Krishnamurthy, R.~Willett, Radioactive scandium in the youngest galactic
  supernova remnant g1. 9+ 0.3, The Astrophysical Journal Letters 724~(2)
  (2010) L161.

\bibitem{leblanc2008scandium}
F.~Leblanc, G.~Alecian, Scandium: a key element for understanding am stars,
  Astronomy \& Astrophysics 477~(1) (2008) 243--247.

\bibitem{do2018super}
T.~Do, W.~Kerzendorf, Q.~Konopacky, J.~M. Marcinik, A.~Ghez, J.~R. Lu, M.~R.
  Morris, Super-solar metallicity stars in the galactic center nuclear star
  cluster: Unusual sc, v, and y abundances, The Astrophysical Journal Letters
  855~(1) (2018) L5.

\bibitem{rojo2013ground}
P.~Rojo, et~al., Ground-based detection of calcium and possibly scandium and
  hydrogen in the atmosphere of hd209458b., Astronomy \&
  Astrophysics/Astronomie et Astrophysique 557~(1).

\bibitem{zhang2008non}
H.~Zhang, T.~Gehren, G.~Zhao, A non-local thermodynamic equilibrum study of
  scandium in the sun, Astronomy \& Astrophysics 481~(2) (2008) 489--497.

\bibitem{rice1997impurity}
J.~Rice, J.~Terry, J.~Goetz, Y.~Wang, E.~Marmar, M.~Greenwald, I.~Hutchinson,
  Y.~Takase, S.~Wolfe, H.~Ohkawa, et~al., Impurity transport in alcator c-mod
  plasmas, Physics of Plasmas 4~(5) (1997) 1605--1609.

\bibitem{1996A&A...310..872A}
G.~{Alecian}, {Fm-Am stars in open clusters as a tool for stellar physics.},
  \aap 310 (1996) 872--878.

\bibitem{rice1995x}
J.~Rice, M.~Graf, J.~Terry, E.~Marmar, K.~Giesing, F.~Bombarda, X-ray
  observations of helium-like scandium from the alcator c-mod tokamak, Journal
  of Physics B: Atomic, Molecular and Optical Physics 28~(5) (1995) 893.

\bibitem{jofre2015gaia}
P.~Jofr{\'e}, U.~Heiter, C.~Soubiran, S.~Blanco-Cuaresma, T.~Masseron,
  T.~Nordlander, L.~Chemin, C.~Worley, S.~Van~Eck, A.~Hourihane, et~al., Gaia
  fgk benchmark stars: abundances of $\alpha$ and iron-peak elements, Astronomy
  \& astrophysics 582 (2015) A81.

\bibitem{singh2020revised}
G.~Singh, A.~Singh, T.~Nandi, Revised and extended calculations with
  relativistic and qed corrections of selected he-like 3d-elemental ions having
  astrophysical significance, Radiation Physics and Chemistry 172 (2020)
  108866.

\bibitem{jonsson2013new}
P.~J{\"o}nsson, G.~Gaigalas, J.~Biero{\'n}, C.~F. Fischer, I.~Grant, New
  version: Grasp2k relativistic atomic structure package, Computer Physics
  Communications 184~(9) (2013) 2197--2203.

\bibitem{palmeri2012atomic}
P.~Palmeri, P.~Quinet, C.~Mendoza, M.~Bautista, J.~Garc{\'\i}a, M.~Witthoeft,
  T.~Kallman, Atomic decay data for modeling k lines of iron peak and light
  odd-z elements, Astronomy \& Astrophysics 543 (2012) A44.

\bibitem{cowan1981theory}
R.~D. Cowan, The theory of atomic structure and spectra, no.~3, Univ of
  California Press, 1981.

\bibitem{indelicato1988multiconfiguration}
P.~Indelicato, Multiconfiguration dirac-fock calculations of transition
  energies in two electron ions with z: 10- 92, Nuclear Instruments and Methods
  in Physics Research Section B: Beam Interactions with Materials and Atoms
  31~(1-2) (1988) 14--20.

\bibitem{grant2007relativistic}
I.~P. Grant, Relativistic quantum theory of atoms and molecules: theory and
  computation, Vol.~40, Springer Science \& Business Media, 2007.

\bibitem{bo2008theoretical}
Q.~Bo, C.~Shao-Hao, G.~Xiang, L.~Jia-Ming, Theoretical study of interesting
  fine-structure splittings for 23p0, 1, 2 states along helium isoelectronic
  sequence, Chinese Physics Letters 25~(7) (2008) 2448.

\bibitem{drake1988theoretical}
G.~Drake, Theoretical energies for the n= 1 and 2 states of the helium
  isoelectronic sequence up to z= 100, Canadian Journal of Physics 66~(7)
  (1988) 586--611.

\bibitem{plante1994relativistic}
D.~Plante, W.~Johnson, J.~Sapirstein, Relativistic all-order many-body
  calculations of the n= 1 and n= 2 states of heliumlike ions, Physical Review
  A 49~(5) (1994) 3519.

\bibitem{natarajan2008kbeta}
A.~Natarajan, L.~Natarajan, K$\beta$ x-ray emission from he-like ions, Journal
  of Quantitative Spectroscopy and Radiative Transfer 109~(12-13) (2008)
  2281--2290.

\bibitem{glowacki2020relativistic}
L.~G{\l}owacki, Relativistic configuration interaction calculations of
  transitions for low-lying states in the helium isoelectronic sequence, Atomic
  Data and Nuclear Data Tables 133 (2020) 101344.

\bibitem{yerokhin2019theoretical}
V.~Yerokhin, A.~Surzhykov, Theoretical energy levels of 1 sns and 1 snp states
  of helium-like ions, Journal of Physical and Chemical Reference Data 48~(3)
  (2019) 033104.

\bibitem{si2016energy}
R.~Si, X.~Guo, K.~Wang, S.~Li, J.~Yan, C.~Chen, T.~Brage, Y.~Zou, Energy levels
  and transition rates for helium-like ions with z= 10--36, Astronomy \&
  Astrophysics 592 (2016) A141.

\bibitem{gu2008flexible}
M.~F. Gu, The flexible atomic code, Canadian Journal of Physics 86~(5) (2008)
  675--689.

\bibitem{aggarwal2012energy}
K.~M. Aggarwal, F.~P. Keenan, Energy levels, radiative rates and electron
  impact excitation rates for transitions in he-like cl xvi, k xviii, ca xix
  and sc xx, Physica Scripta 85~(2) (2012) 025306.

\bibitem{grant1980atomic}
I.~Grant, B.~McKenzie, P.~Norrington, D.~Mayers, N.~Pyper, An atomic
  multiconfigurational dirac-fock package, Computer Physics Communications
  21~(2) (1980) 207--231.

\bibitem{massacrier2012extensive}
G.~Massacrier, M.-C. Artru, Extensive spectroscopic data for multiply ionized
  scandium: Sc iii to sc xxi, Astronomy \& Astrophysics 538 (2012) A52.

\bibitem{bhatti2001mcdf}
M.~I. Bhatti, M.~Bucardo, W.~F. Perger, Mcdf calculations of the specific mass
  shift in helium-like ions, Journal of Physics B: Atomic, Molecular and
  Optical Physics 34~(3) (2001) 223.

\bibitem{boiko1978x}
V.~Boiko, A.~Y. Faenov, S.~Pikuz, X-ray spectroscopy of multiply-charged ions
  from laser plasmas, Journal of Quantitative Spectroscopy and Radiative
  Transfer 19~(1) (1978) 11--50.

\bibitem{beiersdorfer1989experimental}
P.~Beiersdorfer, M.~Bitter, S.~Von~Goeler, K.~Hill, Experimental study of the
  x-ray transitions in the heliumlike isoelectronic sequence, Physical Review A
  40~(1) (1989) 150.

\bibitem{hill1979determination}
K.~Hill, S.~Von~Goeler, M.~Bitter, L.~Campbell, R.~Cowan, B.~Fraenkel,
  A.~Greenberger, R.~Horton, J.~Hovey, W.~Roney, et~al., Determination of fe
  charge-state distributions in the princeton large torus by bragg crystal
  x-ray spectroscopy, Physical Review A 19~(4) (1979) 1770.

\bibitem{NIST}
Nist database, \url{http://https://www.nist.gov/pml/atomic-spectra-database}.

\bibitem{sugar1985atomic}
J.~Sugar, C.~Corliss, Atomic energy levels of the iron-period elements:
  potassium through nickel, Tech. rep., American Chemical Society (1985).

\bibitem{kaufman1988wavelengths}
V.~Kaufman, J.~Sugar, Wavelengths and energy level classifications of scandium
  spectra for all stages of ionization, Journal of physical and chemical
  reference data 17~(4) (1988) 1679--1789.

\bibitem{martin1988atomic}
G.~Martin, J.~Fuhr, W.~L. Wiese, Atomic transition probabilities scandium
  through manganese, Journal of Physical and Chemical Reference Data,
  Supplement 17~(3) (1988) 1--512.

\bibitem{Anu:2013}
R.~L. Kurucz, \url{http://kurucz.harvard.edu}.

\bibitem{fischer2019grasp2018}
C.~F. Fischer, G.~Gaigalas, P.~J{\"o}nsson, J.~Biero{\'n}, Grasp2018—a
  fortran 95 version of the general relativistic atomic structure package,
  Computer Physics Communications 237 (2019) 184--187.

\bibitem{ekman2019ris}
J.~Ekman, P.~J{\"o}nsson, M.~Godefroid, C.~Naz{\'e}, G.~Gaigalas,
  J.~Biero{\'n}, Ris 4: a program for relativistic isotope shift calculations,
  Computer Physics Communications 235 (2019) 433--446.

\bibitem{mann1971breit}
J.~B. Mann, W.~R. Johnson, Breit interaction in multielectron atoms, Physical
  Review A 4~(1) (1971) 41.

\bibitem{mackenzie1980program}
B.~Mackenzie, I.~Grant, P.~Norrington, A program to calculate transverse breit
  and qed corrections to energy levels in a multiconfiguration dirac-fock
  environment, Computer Physics Communications 21~(2) (1980) 233--246.

\bibitem{grant1974gauge}
I.~Grant, Gauge invariance and relativistic radiative transitions, Journal of
  Physics B: Atomic and Molecular Physics 7~(12) (1974) 1458.

\bibitem{gaigalas2001program}
G.~Gaigalas, S.~Fritzsche, I.~P. Grant, Program to calculate pure angular
  momentum coefficients in jj-coupling, Computer physics communications 139~(3)
  (2001) 263--278.

\bibitem{osti_4571333}
L.~J. Armstrong, \href{https://www.osti.gov/biblio/4571333}{Theory of the
  hyperfine structure of free atoms.}
\newline\urlprefix\url{https://www.osti.gov/biblio/4571333}

\bibitem{jonsson1996hfs92}
P.~J{\"o}nsson, F.~A. Parpia, C.~F. Fischer, Hfs92: A program for relativistic
  atomic hyperfine structure calculations, Computer physics communications
  96~(2-3) (1996) 301--310.

\bibitem{verdebout2014hyperfine}
S.~Verdebout, C.~Naze, P.~J{\"o}nsson, P.~Rynkun, M.~Godefroid, G.~Gaigalas,
  Hyperfine structures and land{\'e} gj-factors for n= 2 states in beryllium-,
  boron-, carbon-, and nitrogen-like ions from relativistic configuration
  interaction calculations, Atomic Data and Nuclear Data Tables 100~(5) (2014)
  1111--1155.

\bibitem{blundell1987reformulation}
S.~Blundell, P.~Baird, C.~Palmer, D.~Stacey, G.~Woodgate, A reformulation of
  the theory of field isotope shift in atoms, Journal of Physics B: Atomic and
  Molecular Physics 20~(15) (1987) 3663.

\bibitem{torbohm1985state}
G.~Torbohm, B.~Fricke, A.~Ros{\'e}n, State-dependent volume isotope shifts of
  low-lying states of group-iia and-iib elements, Physical Review A 31~(4)
  (1985) 2038.

\bibitem{shabaev1994relativistic}
V.~Shabaev, A.~Artemyev, Relativistic nuclear recoil corrections to the energy
  levels of multicharged ions, Journal of Physics B: Atomic, Molecular and
  Optical Physics 27~(7) (1994) 1307.

\bibitem{dyall1989grasp}
K.~Dyall, I.~Grant, C.~Johnson, F.~Parpia, E.~Plummer, Grasp: A general-purpose
  relativistic atomic structure program, computer physics communications 55~(3)
  (1989) 425--456.

\bibitem{stone2005table}
N.~Stone, Table of nuclear magnetic dipole and electric quadrupole moments,
  Atomic Data and Nuclear Data Tables 90~(1) (2005) 75--176.

\bibitem{accad1971s}
Y.~Accad, C.~Pekeris, B.~Schiff, S and p states of the helium isoelectronic
  sequence up to z= 10, Physical review A 4~(2) (1971) 516.

\end{thebibliography}
\end{document}